\documentclass[doublecol,figures]{epl2}
\usepackage{amssymb}
\title{Na atomic order, Co charge disproportionation and magnetism in
Na$_{x}$CoO$_{2} $ for large Na contents}%
\shorttitle{Na atomic order, Co charge disproportionation and magnetism in
Na$_{x}$CoO$_{2} $ for large Na contents}

\author{H.~Alloul\inst{1}\thanks{E-mail: \email{alloul@lps.u-psud.fr}}
\and I.R.~Mukhamedshin\inst{1,2} \and G.~Collin\inst{3} \and
N.~Blanchard\inst{1}} \shortauthor{H.~Alloul \etal}

\institute{
  \inst{1} Laboratoie de Physique des Solides, UMR CNRS 8502, Univ. Paris-Sud, 91405 Orsay, France\\
  \inst{2} Physics Department, Kazan State University, 420008 Kazan, Russia\\
  \inst{3} Laboratoire L\'eon Brillouin, CE Saclay, CEA-CNRS, 91191 Gif-sur-Yvette, France
}


\abstract{We have synthesized and characterized four different stable phases
of Na ordered Na$_{x}$CoO$_{2}$, for $0.65<x<0.8$. Above 100~K they display
similar Curie-Weiss susceptibilities as well as ferromagnetic $q=0$ spin
fluctuations in the CoO$_{2}$ planes revealed by $^{23}$Na NMR data. In all
phases from $^{59}$Co NMR data we display evidences that the Co
disproportionate already above 300~K into non magnetic Co$^{3+}$ and
magnetic $\approx $Co$^{3.5+}$ sites on which holes delocalize. This allows
us to understand that metallic magnetism is favored for these large Na
contents. Below 100~K the phases differentiate, and a magnetic order sets in
only for $x\gtrsim 0.75$ at $T_{N}=$22~K. We suggest that the charge order
also governs the low $T$ energy scales and transverse couplings.}

\begin{document}

\maketitle

\section{Introduction}

The interplay between magnetism and metallicity in the copper oxygen plane
structure of the cuprates is at the origin of the great interest which has
been devoted to their high temperature superconductivity. Such an interplay
has been also found recently in the lamellar cobaltates Na$_{x}$CoO$_{2}$,
which further display a high thermoelectric power together with good
electronic conductivity \cite{Terasaki}. In both systems metallicity is
induced by doping the square CuO$_{2}$ or triangular CoO$_{2}$ layers by
insertion of dopants in the ionic separation layers. Soon after the
discovery of the cuprate superconductivity, it has been suggested that the
electron gas could display intrinsic inhomogeneous charge structures, such
as linear ''stripes'' of Cu$^{2+}$ spins separated by doped metallic stripes
\cite{Emery}. However, such static structures have only been found in
specific cases \cite{Tranquada}, and the magnetic properties of the Cu sites
have been usually found rather homogeneous.

A distinct property of the Co ions in the large crystal field induced by
their oxygen octahedral environment in the CoO$_{2}$ structure, is that the $%
t_{2g}$ triplet of the Co site is much lower in energy than the $e_{g}$
doublet, so that the electronic structure of the Co ions is expected to
correspond to low spin configurations Co$^{3+}$ ($S=0$) or Co$^{4+}$ ($S=1/2$%
) obtained by filling only the $t_{2g}$ triplet states. Consequently ordered
charge structures are more frequently expected than for cuprates. Indeed,
using the quasi-unique site sensitivity of NMR techniques, a Na atomic order
\cite{NaPaper} and an associated Co ordered charge disproportionation (OCD)
\cite{CoPaper} has been revealed for a specific metallic $x_{0}\approx 0.7$
composition, which displays a Curie-Weiss susceptibility. However for the
metallic cobaltate antiferromagnetic (AF) phases found for $x\geqslant 0.75$
\cite{Sugiyama,Mendels}, neutron scattering experiments have been analyzed
in a uniform local moment picture \cite{Boothroyd,Bayrakci}. This is
unexpected for hole doping of Na$_{1}$CoO$_{2}$ which is a band insulator
built from the non magnetic Co$^{3+}$ state \cite{Lang}, for which the $%
t_{2g}$ multiplet is filled by the six $d$ electrons.

Here we present the first systematic effort undertaken so far to study with
local probes the variation with $x$ of the magnetic properties and
demonstrate that the OCD is generic for $x>0.65$ including the AF phase with
$T_{N}=$22~K. This OCD always results into non magnetic Co$^{3+}$ and a
second type of sites with formal valence of about 3.5 responsible for the
peculiar magnetic properties. We establish that the latter display above
100~K a generic nearly ferromagnetic behaviour for four phases displaying
distinct ordered Na structures, so that the OCD is an essential ingredient
to explain the magnetic properties of these high $x$ phases. Surprisingly
the low $T$ ground state magnetic properties differ markedly as, apart the
AF phase, the former $x_{0}\approx 0.7$ phase appears an experimental
realization of a nearly ferromagnetic 2-dimensional metal without static
magnetic order down to $T=0$. The hole doped AF phases have been shown
ferromagnetic in plane and AF between planes (A-type AF) \cite
{Boothroyd,Bayrakci}. So, the Na atomic order and the OCD of Co appear
essential as well in governing the low energy modifications of the band
structure of the correlated metallic state, which drive the coupling between
Co planes and the low $T$ magnetic and thermoelectric properties \cite{Ong}.

\begin{figure}[tbp]
\onefigure[width=1\linewidth]{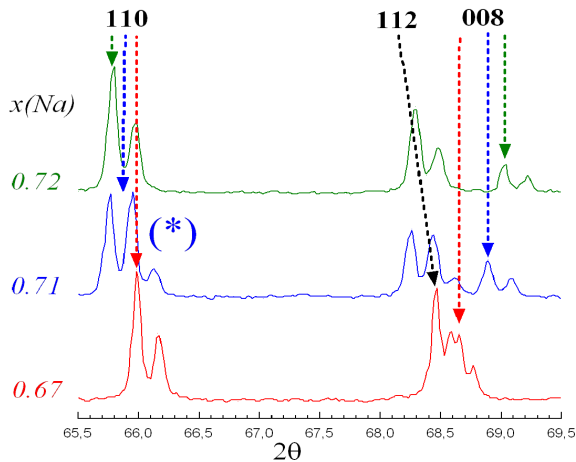}%
\caption{Hexagonal (P63/mmc N$^\circ$ 196) indexation of part of the X ray
diffraction patterns. At these high angle values the Cu K$\alpha $1 and
K$\alpha $2 splits the reflections into two separate peaks with intensity
ratio 2/1. (*) In the 0.71 case an additional structural splitting separates
the hexagonal 110 (and 112) reflections into two distinct 310 and
020 orthorhombic (Ccmm N$^\circ$ 63) peaks.}%
\label{FigXray}
\end{figure}

\section{Samples}

We have synthesized our samples by standard solid state reaction of Na$_{2}$
CO$_{3}$ and Co$_{3}$O$_{4}$ powders in flowing oxygen, with nominal
concentrations $x$ increasing by increments of 2 to 3/1000 in the range
$0.65\leqslant x\lesssim 0.80$. X ray powder diffraction data always
exhibited the Bragg peaks corresponding to the two-layer Co structure with a
hexagonal unit cell. However we systematically detected weak additional
reflections indicative of three dimensional Na long range ordering. It has
been immediately clear that the corresponding Na order is complicated and
highly dependent on Na content. This ordering was even found for a specific
composition to drive an orthorhombic distortion of the average lattice as
shown for $x=0.71$ in fig.~\ref{FigXray}, on a detail of the Bragg peaks of
the X ray diffraction profile. From the Rietveld refinements of the X ray
data summarized in table~\ref{table1} we could isolate four distinct phases
for $0.65\leqslant x\lesssim 0.80$. Each phase exhibits a specific Na
ordering leading to characteristic additional diffractions: commensurate
reflections, or incommensurate superstructure satellites with $q(b^{\ast})$
as the component of the wave vector modulation.

The concentrations for which single phase samples could be stabilized are
sequenced in four distinct narrow $x$ non overlapping domains outside which
the synthesized samples were found as intermediate mixtures of these phases.
In addition, multiphasing has been found to occur quite commonly for any $x$
value if no particular care to homogeneize the samples was taken. We could
however synthesize these four phases reproducibly, forbidding any air
exposure of the samples. Confirmation that nearly pure phases could be
achieved on large mass samples ($\simeq $400~mg) required for NMR
measurements has been directly obtained from $^{23}$Na and $^{59}$Co NMR
data, as shown hereafter.

The orderings found for these four phases always correspond to a symmetry
lowering and is systematically based on an orthorhombic reference subcell
(Ccmm, N$^{\circ }$63): $a_{ort}=a_{hex}\sqrt{3}$; $b_{ort}=a_{hex}$; $%
c_{ort}=c_{hex}$. Owing to the orthorhombic distortion for $x=0.71$ the
labelling H67, O71, H72 and H75 has been used hereafter and in table~\ref
{table1}. The formerly studied $x_{0}\approx 0.7$ phase \cite
{NaPaper,CoPaper} is in fact that with the lowest $x$ value $x=0.67(1)$. We
indeed found that any sample with larger $x$ evolves towards this limiting
composition if it is kept in insufficiently dry atmosphere.

\begin{largetable}%
\caption{Parameters of the studied phases. The accuracy on their difference
of Na content is much better than that on $x$ itself ($\pm $ 0.01). The
lattice constants $a,b,c$ corresponding to fig.~\ref{FigXray} are given in
the hexagonal or orthorhombic reference cell. When determined, the
incommensurate $q(b^{\ast})$ or commensurate modulations $q(c^{\ast})$ are
given. The Co$^{3+}$ fraction $y$ is obtained from $^{59}$Co NMR intensity
data. The 0.67 sample displays a commensurate
orthorombic superstructure $(a_{hex}\sqrt{3},\;3a_{hex},\;3c_{hex})$.}%
\label{table1}.
\begin{center}
\begin{tabular}{cccccccccc}
Phase & $x$ & $a_{ort}/\surd{3}$ (\AA) & $a_{hex}$ or $b_{ort}$ (\AA) & $c$ (\AA) & $q(b^{\ast})$ & $q(c^{\ast})$ & $A_{eff}^{iso}$ (kG/$\mu _{B}$)& y (\%) & \\
\hline
H67 & $0.67$ & $a_{hex}$ & 2.82920(1) & 10.9387(4) & $1/3$ & $1/3$ & 9.1(3) & 26(4) &  \\
O71 & $0.71$ & 2.83931(2) & 2.83031(3) & 10.8929(2) & 0.2849(1) & 0 & 8.0(3) & 40(5) &  \\
H72 & $0.72$ & $a_{hex}$ & 2.83651(4) & 10.8770(2) & 0.2810(1) & 0 & 7.3(3) & 37(5) &  \\
H75 & $\geq 0.75$ & $a_{hex}$ & 2.84162(1) & 10.8058(3) & - & - & 7.8(3) & 33(4) & \\
\end{tabular}
\end{center}
\end{largetable}

\section{Magnetic susceptibility and $^{23}$Na NMR shifts.}

\begin{figure}[tbp]
\onefigure[width=1\linewidth]{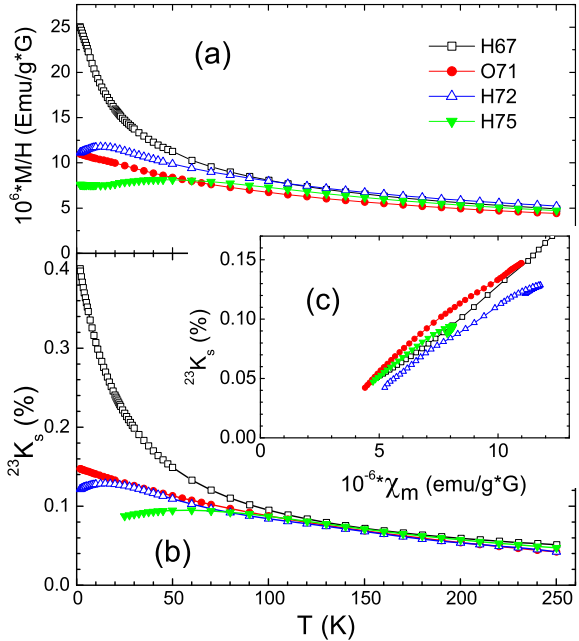}%
\caption{(a) $T$-dependencies of the bulk susceptibility $\chi_{m}$ measured
with a DC SQUID in a 5~T field; (b) $T$ dependence of the mean $^{23}$Na NMR
shift. Identical behaviour above 100~K can be seen for the four phases with
remarkable differences at low $T$; (c) the linear variations of $K_{s}$
versus $\chi_{m}$ underlines the phase purity of the samples. The H67 data
for $T<$30~K \cite{NaPaper} has been omitted to better display
those of the new phases.}%
\label{fig.2}
\end{figure}

The single crystal grains of samples of these phases were oriented in the $%
H_{0}=7$ Tesla NMR field within Stycast or paraffin. SQUID measurements of
the macroscopic susceptibility $\chi _{m}$ taken in 5~T field allow us to
evidence that the different phases display different magnetic properties.
For instance, as evidenced in fig.~\ref{fig.2}a, the low T magnitude of
$\chi_{m}$ decreases progressively with increasing $\emph{x}$. The H75 phase
is furthermore found to be the only phase in which a magnetic order is
detected in low applied field. However, as minute amounts of impurities or
slight admixture of phases could spoil the bulk measurements, spectroscopic
measurements with local probes better determine the susceptibility of each
phase.

\begin{figure}[tbp]
\onefigure[width=1\linewidth]{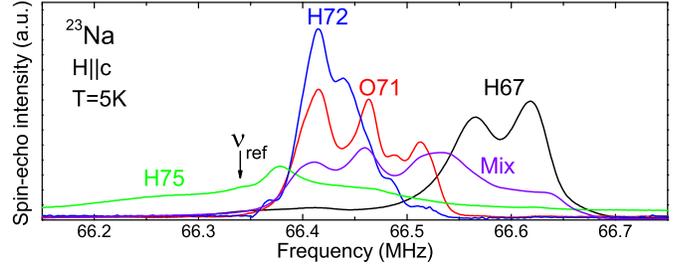}%
\caption{$^{23}$Na NMR central line spectra taken at 5~K. They are quite
distinct for the four nearly pure phases, with some overlap between O71 and
H72 spectra. That for a sample which is clearly a mixture of H67, O71 and
H72 is shown as well. $\nu_{ref}$ is the non magnetic $^{23}$Na NMR
reference.}%
\label{fig.1}
\end{figure}

As reported in ref.~\cite{NaPaper} on H67, the $^{23}$Na NMR is an excellent
probe of both Na order and of the magnetic properties. Indeed the $^{23}$Na
NMR displays distinct quadrupole splittings for the different Na sites of
this structure. In view of the more complex structures of the new phases
there was no surprise in finding a larger number of resolved Na sites,
although with similar magnitudes of their quadrupole splittings. The
magnetic properties of the compounds are probed at the local scale through
the NMR shifts of the different Na sites resolved in the $(-\frac{1}{2}%
\leftrightarrow \frac{1}{2})$ transition of the $^{23}$Na spectra presented
in fig.~\ref{fig.1}. There one can see that the $T=$5~K spectra are quite
distinct for the four phases. For H75 a large broadening occurs in the AF
state below $T_{N}=22$~K, while the spectrum of H67 is much more shifted
than those of O71 and H72, which are distinct but display some overlap. The
H72 batch has been found to evolve in time at room $T$ towards O71, and a
slight mixture of the two pure phases might be unavoidable in the bulk
samples \footnote{In various published papers for x values in the range
studied, e.g. \cite{Sakurai}, we could find on the susceptibility data many
signs indicating that the samples were superposition of our phases.}. Quite
generally $^{23}$Na NMR allowed us to control the phase purity of the bulk
of the NMR sample, as multiphase samples display superimposed spectra, as
seen in fig.~\ref{fig.1}.

Let us recall, as detailed in ref.~\cite{NaPaper} that for a field $%
H_{0}\parallel \alpha $, the NMR shift $K_{\beta}^{\alpha}$ of a Na atomic
site $\beta$ probes the spin susceptibility $\chi _{s,i}^{\alpha }(T)$ of
the neighbouring Co sites $i$ through transferred hyperfine couplings $%
A_{\beta ,i}^{\alpha }$ with $K_{\beta }^{\alpha }=\sum_{i}A_{\beta
,i}^{\alpha }\;\chi _{s,i}^{\alpha }(T)$. The main result found for H67, and
verified as well here for the other phases, is that the $K_{\beta }^{\alpha
}(T)$ variations scale with eachother for all Na sites. This $T$ dependence
is associated with the average $\chi _{s}^{\alpha }(T)$ of the magnetic Co
sites of the structure. So, overlooking the diversity of Na sites, the first
moment (or center of gravity) of the $^{23}$Na NMR signal writes $%
K_{s}^{\alpha }=A_{eff}^{\alpha }\chi _{s}^{\alpha }(T)$ where $%
A_{eff}^{\alpha }$ is an effective hyperfine field per Co site. The $T$
variations of $K_{s}^{\alpha }$ are reported in fig.~\ref{fig.2}b, and are
shown to be quite identical for $T>100$~K with a unique Curie-Weiss
$(T-\Theta)^{-1}$ variation (with $\Theta\approx$-80~K). They differ
markedly below 100~K, as does the SQUID data for $\chi _{m}$, the low $T$
enhancement of $\chi _{s}^{\alpha }(T)$ observed for H67 being progressively
reduced for increasing $x$.

The usual comparison between the SQUID and Na NMR data, displayed in the
good linear $K_{s}$ versus $\chi _{m}$ plots of fig.~\ref{fig.2}c allows us
to confirm the purity of the isolated phases. The high-T slopes of these
plots yield similar values for the effective hyperfine coupling
$A_{eff}^{iso}$ (table~\ref{table1}), which could be expected as $^{23}$Na
sites are coupled with many cobalts \cite{NaPaper}. In all phases the
anisotropy of $\chi _{s}^{\alpha }(T)$, given by that of $K_{s}^{\alpha }$,
has been found $\lesssim $ $\pm $ 0.1.

In the H75 AF phase, the saturation of $K_{s}(T)$, that is $\chi
_{s}^{\alpha }(T)$, seen at low $T$ in fig.~\ref{fig.2}b should be
associated with the onset of AF correlations. In a uniform Heisenberg model,
one would then assign the progressive increase of $\chi _{s}^{\alpha }(T)$
at low $T$ with decreasing $x$ to a decrease of $T_{N}$ and of out of plane
AF coupling strength. However this primary interpretation fails as NMR data
taken down to 1.4~K (and $\mu $SR to 50~mK \cite{Mendels2}), did not
evidence any frozen magnetic state in the three other phases, which are then
\textit{paramagnets in their ground state}, most probably metallic, as no
electronic magnetic transition is detected.

\begin{figure}[tbp]
\onefigure[width=1\linewidth]{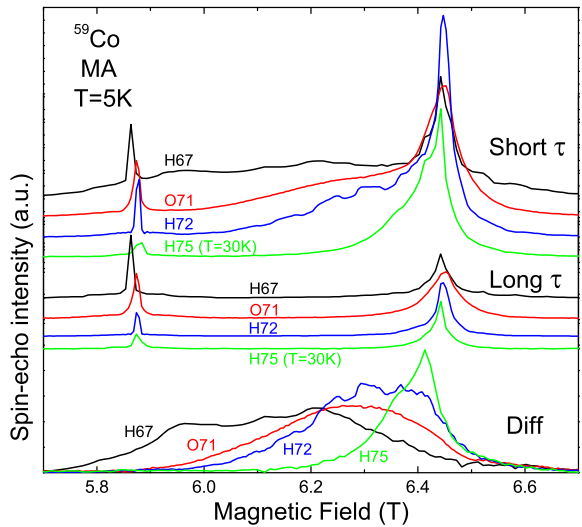}%
\caption{Spectra taken with a large pulse spacing $\tau =200~\mu s$ in a
spin echo sequence allowed us to isolate the narrow spectra of Co1 sites
with long $T_{2}$. The broad spectra of the magnetic Co2 sites with short
$T_{2}$ are obtained by subtracting the Co1 spectra from those taken with
$\tau =10~\mu s$. The average shift of the Co2 sites decreases with
increasing $x$ as does the $^{23}$Na shift in figs.~\ref
{fig.2}b and \ref{fig.1}.}%
\label{fig.3}
\end{figure}

\section{$^{59}$Co NMR}

The difference between phases appears as well in the $^{59}$Co NMR spectra,
which also display more Co sites than for H67 \cite{CoPaper}.

\subsection{Charge disproportionation}

As in H67 we identified two classes of Co sites, their $(-\frac{1}{2}%
\rightarrow \frac{1}{2})$ transitions being easily visualized when $H_{0}$
is applied at 54.7$^{\circ }$ with respect to the $c$ axis, the ''magic
angle'' for which quadrupole effects are reduced, as seen in
fig.~\ref{fig.3}. A first series which we label as the Co1 ''class'', is
associated with non magnetic Co$^{3+}$, with small spin lattice $T_{1}^{-1}$
and spin spin $T_{2}^{-1}$ relaxation rates, and occurs with similar NMR
shifts at low $T$ in all four phases. The more complex spectra of fast
relaxing magnetic Co sites, which we label as the Co2 ''class'' is seen in
fig.~\ref{fig.3} to include diverse sites with much larger shifts at low $T$
than the Co1 sites. The increase of average shift of these Co2 with
decreasing $x$ agrees perfectly with the SQUID and $^{23}$Na NMR data
(fig.~\ref{fig.2}).

\begin{figure}[tbp]
\onefigure[width=1\linewidth]{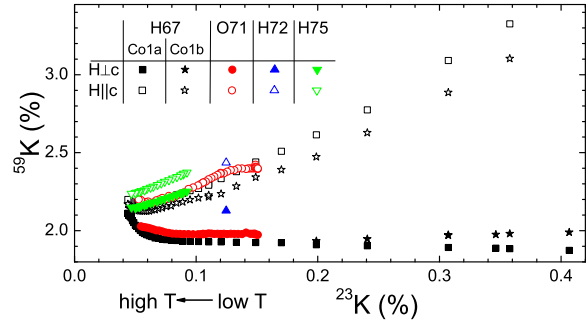}%
\caption{At low $T$, the Co1 sites NMR shifts display linear variations
versus the $^{23}$Na shifts $^{23}K$. For $H_{0}\parallel c$, the Co1
hyperfine coupling\ with the magnetic Co2 sites have similar magnitudes for
all samples. For $H_{0}\perp c$ it has a similar magnitude for H72 and H75
but nearly vanishes for H67 and O71. For H67 two Co1 sites are well
resolved, and their shifts depart slightly below 40~K (i.e. $^{23}K>0.2\%$),
as reported in ref.~\cite{Gavilano}. The upturns for small $^{23}K$ are due
to the onset of Co1-Co2 site exchange due to Na motion, as discussed in the
text.}%
\label{fig.5}
\end{figure}

The Co1 NMR shifts were found to vary with temperature, which is a sign that
these non magnetic sites are sensing the magnetism of the Co2 sites through
transferred hyperfine couplings. Indeed, in Fig.~\ref{fig.5}, we evidence
that the shifts $^{59}K_{1}^{\alpha }$ of the Co1 nuclei scale linearly at
low $T$ with that of $^{23}K$. The  $T$ independent orbital contribution to
$^{59}K_{1}^{\alpha }$ can be obtained by extrapolating this linear
dependence to $^{23}K=0$ (that is vanishing spin susceptibility $%
\chi _{s}(T)$).\ It is found isotropic and increases slightly from 1.95 to
2.05\% from H67 to H75. These values are quite comparable with the purely
orbital shift of 1.95\% found for Co$^{3+}$ in the band insulator Na$_{1}$CoO%
$_{2}$ \cite{Lang}. The fraction $y$ of Co$^{3+}$ sites, estimated from the
Co1 relative NMR intensity (corrected for $T_{2}$ decay), increases
slightly, but not regularly with $x$ (table~\ref{table1}), this overall
trend being expected as all Co sites become Co$^{3+}$ for $x=1$. However as
we find $y<x$, the average valence of the Co2 class of sites is always $<4+$%
, and the OCD detected in H67 is present in all phases, \textit{including
the AF ordered phase}.

\subsection{Na motion}

As seen in Fig.~\ref{fig.5} we could detect the Co1 sites up to room $T$ in
most phases, well above the onset of Na motion detected hereafter at $%
\approx $200~K from $^{23}$Na $T_{1}$ data \cite{MotionPaper}. This implies
that the OCD occurs already above room $T$, contrary to the proposal of
ref.~ \cite{Gavilano}. However in fig.~\ref{fig.5} significant increases of
$^{59}K_{1}^{\alpha }$ with respect to the low $T$ linear $^{23}K$
dependence are observed for $T\gtrsim 150$~K, that is for $^{23}K<0.08$. We
attribute this to the onset of Co1-Co2 site exchange which becomes
significant when Na motion begins to set in. Such site exchanges are usually
easily detected in NMR spectra as they yield a progressive reduction of the
line splitting between the two sites until they merge in a single exchange
narrowed line at very high $T$, when the exchange rate exceeds the line
splitting. Here the increase of the Co1 NMR shift merely corresponds to the
onset of Co1-Co2 site exchange. Indeed from ref.~\cite{CoPaper},
$K^{ab}$(Co2) is as large as $\approx $4\% above 200~K, so the tiny
0.1\%-0.2\% increase of $K^{ab}$(Co1) corresponds to a $\approx $10\%
decrease of the 1.5~MHz Co1-Co2 splitting, that is a very small exchange
rate $\tau _{ex}^{-1}\approx 150$~kHz. This analysis is validated by the
fact that weaker upturns of $K^{c}$(Co1) are seen for $H_{0}\parallel c$ for
which $K^{c}$(Co1) and $K^{c}$(Co2) happen to differ only slightly as seen
in ref.~\cite{CoPaper}.

Therefore this key observation that the rate of exchange $\tau _{ex}^{-1}$
of Co sites is very slow with respect to expectations for electronic
processes allows us to conclude that the Co1-Co2 site exchange is connected
with Na motion. This proves that \textit{the Co charge does correlate with
the Na environment} (e.g. Na1 sites being on top of Co$^{3+}$).

\section{$\chem{^{23}}$Na spin lattice relaxation and electron spin dynamics}

\begin{figure}[tbp]
\onefigure[width=1\linewidth]{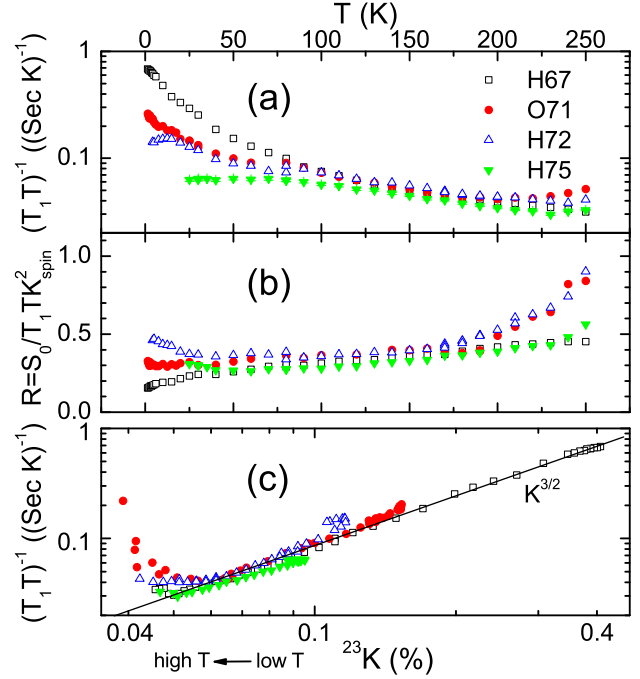}%
\caption{$T$ variation of
$(T_{1}T)^{-1}$ (a) and the normalized Korringa product $R$ (b) for the four
phases. While the data are distinct below 100~K, in (c) a universal scaling
between $(T_{1}T)^{-1}$ and $^{23}K$ is shown to apply. The high $T$
deviations due to Na motion and the slight low $T$ increases for H72 and O71
are discussed in the text.}%
\label{fig.4}
\end{figure}

To search for differences in the dynamic electronic susceptibilities of
these phases we have taken extensive $^{23}$Na $T_{1}$ data. As $^{23}$Na
has a spin $I=3/2$, its nuclear magnetization recovery should be given by

\begin{equation}
M(t)/M_{0}=1-W\exp (-6t/T_{1})-(1-W)\exp (-t/T_{1}),
\end{equation}

with $W=0.9$ if only the central transition has been saturated. Since
condition being impossible to fulfill strictly experimentally, $W$ has been
left as an adjustable parameter, which was found to evolve between 0.9 and
0.7 depending on the sample and experimental conditions. The $T_{1}^{-1}$
data were found slightly anisotropic, i.e. $\approx $30\% larger for
$H_{0}\perp c$ than for $H_{0}\parallel c$, for $T<$200~K. So, in fig.~\ref
{fig.4} we only plotted the data for $H_{0}\parallel c$.

Although the low $T$ variations of $\left( T_{1}T\right)^{-1}$ differ for
the four phases, they do become identical above 100~K as seen in fig.~\ref
{fig.4}a. There, we assign the extra high $T$ contribution to the occurrence
of Na motion, their onset taking place at distinct $T\gtrsim 200$~K for the
different phases. As can be anticipated, we could check that the magnetic
behaviour is not that of a Fermi liquid, especially for the H67 sample, as
$\left(T_{1}T\right)^{-1}$ and $^{23}K(T)$ do not combine into a constant
$R=S_{0}/T_{1}T(^{23}K)^{2}$, where $S_{0}=(\hbar /4\pi k_{B})(\gamma
_{e}/\gamma _{n})^{2}$ is the universal Korringa ratio. In fig.~\ref{fig.4}b
$R$ is of course sample independent between 80 and 160~K and always smaller
than unity, as noticed in a $x\approx 0.7$ mixed phase sample by Ihara
\textit{et al.} \cite{Ishida}. This is expected if a ferromagnetic
quasi-elastic peak at \textbf{q}$\approx 0$ dominates the spin excitations
as revealed by Inelastic Neutron Scattering (INS) for the H75 phase above
$T_{N}$ \cite {Boothroyd}. Indeed, in such a case the \textbf{q}$\approx 0$
response enhances markedly $\chi _{s}($\textbf{q}$=0)$, that is
$^{23}K_{s}$, while $\left( T_{1}T\right) ^{-1}$\ is less enhanced as it
probes $\chi ^{\prime \prime }($\textbf{q}$,\omega)$ at all \textbf{q}
values. The identical behaviour found here above 100~K extends this result
to all our Curie-Weiss phases.

To characterize further the difference of magnetic excitations in the
diverse ground states we searched for relationships between $(T_{1}T)^{-1}$
and $^{23}K$, as done in fig.~\ref{fig.4}c. Spin fluctuation theories in
nearly ferromagnetic metals are expected to give $(T_{1}T)^{-1}=aK^{n}$
\cite{Moriya-Ueda}, with $n=1$ in 3D \cite{Alloul-Mihaly}, while $n=3/2$ is
expected for 2D \cite{Hatatani-Moriya,Kitagawa}. For H67, we do remarkably
find an accurate scaling, with $n=1.5\pm 0.1$, that is $%
(T_{1}T)^{-1}=aK^{3/2}$ over the entire range 1.5~K$<T<$300~K. This
interpretation of the data implies that  the Curie-Weiss temperature
$\Theta$=-80~K found in $K(T)$ and $\chi(T)$ is not associated with AF
in-plane couplings but results from the nearly ferromagnetic electronic band
behaviour \cite{Moriya-Ueda,Hatatani-Moriya} as seen for instance in
TiBe$_{2}$ \cite{Alloul-Mihaly}. Impressively this 2D \textit{ferromagnetic}
scaling extends down to low $T$ with a unique $a$ for all phases \footnote{
For H72 and O71, the small deviations for $T\lesssim 5$K are not intrinsic
as data for the resolved Na lines were not found to scale perfectly.
Purification of these phases is required for accurate low $T$ studies.},
including the AF phase down to $T_{N}$.

At this stage one might wonder whether the saturation of $^{23}K(T)$
observed below 100~K in H75 is related to AF plane to plane couplings. Such
AF fluctuations that enhance $\chi ^{\prime \prime }(\mathbf{q}=\mathbf{q}%
_{AF})$ should result in an increase of $(T_{1}T)^{-1}$ and $R$ with a
divergence at $T_{N}$. But they are not probed by the $^{23}$Na nuclei, as
the local fields induced by two adjacent Co layers cancel in the A type AF
structure, as confirmed by the weak $^{23}$Na NMR shift in the N\'{e}el
state (fig.~\ref{fig.1}). The $^{23}$Na $T_{1}$ only probes then the
strength of the ferro fluctuations, and the perfect scaling found above
proves that the main incidence of AF fluctuations is to reduce the ferro
ones below 100~K. Comparison with the numerical results of Hatatani \emph{et
al.} \cite{Hatatani-Moriya} allows us to point out that, for H67, the low
$T$ increase of $\chi (\mathbf{q}=0)$ with respect to the common Curie-Weiss
variation is that expected by these authors in the immediate vicinity of a
ferromagnetic instability. Therefore the H67 phase \textit{appears as an
ideal 2D nearly ferromagnetic metal without 3D ordering settling in at low}
$T$.

\section{Discussion}

The four phases exhibit specific Na orderings, but similar charge
disproportionation, and identical ferro in plane fluctuations above 100~K,
independent of the details of the OCD. It seems to us that this analogy of
properties relies heavily on the occurrence of non magnetic Co$^{3+}$ as it
reduces the number of hopping paths between magnetic sites with respect to a
homogeneous structure. The associated decrease of bandwidth $W$ and increase
of hole density on the magnetic sites magnify the role of correlations $U$,
which enhances the ferro tendency supported by LDA calculations \cite{Singh}
for the uniform case.

On the contrary the ground state magnetic properties are certainly driven by
the diverse Na atomic orders evidenced here, similar to those suggested, or
observed \cite{Zandbergen,Roger} as well for various other $x$ values. One
could expect then distinct ordered magnetic states as in the case of the $%
x=0.5$ phase for which an AF order inside the Co plane sets in at $T_{N}=$
86~K \cite{Mendels,Gasparovic} in the absence Co$^{3+}$ \cite{Bobroff}. But
for the analogous ferro in plane couplings evidenced here for large $x$, one
would always expect A type AF order to occur. In these AF phases, the 3D
dispersion of the spin wave excitations found by INS has been analyzed with
Heisenberg Co-Co AF coupling between planes either with nearest
\cite{Boothroyd,Bayrakci} or next nearest neighbour exchange through Na
orbitals \cite{Mazin}. For such Heisenberg transverse couplings, a hectic
evolution of $T_{N}$ versus $x$ should presumably result, depending on the
actual Na order, contrary to the smooth evolution of \textit{paramagnetism}
found for $x>0.65$ and the abrupt occurrence of AF above $x=0.75$.

This definitely allows us then to conclude that metallic magnetism is indeed
responsible as well for the low $T$ states for large $x$ in these Na
cobaltates. One might consider \cite{McKenzie} that a Fermi liquid state is
only reached below an energy scale given by the temperature $T^{\ast }$ at
which $\chi _{s}$ saturates, which increases from $\approx $1~K for H67
\cite {NaPaper} to $\approx$ 60~K for H75. The band parameters associated
with Na order would then be responsible for these $T^{\ast }$ values and for
the transverse couplings which drive the AF order. While some attempts have
been done to take into account both correlations and OCD \cite{Marianetti},
extensions to diverse Na atomic orders are required to explain the evolution
with $x$ of the ground state properties \footnote{After the disclosure of an
earlier version of these results \cite{CondMat}, a theoretical attempt to
explain them within a strong correlation approach has been done \cite{Gao}.
The existence of a critical doping value for the onset of in-plane
ferromagnetism has been reproduced, as well as the occurrence of Co$^{3+}$
in the presence of Na disorder, using heuristic values for the Na-Co
interaction potentials.}.

An important experimental aspect revealed by our thorough investigation is
that, contrary to the case of cuprates for which dopant induced disorder is
quite influential, and governs many aspects of the cuprates phase diagram
\cite{FRA-alloul}, the hole doping achieved in bulk cobaltate samples by
insertion of ordered Na planes corresponds to rather clean situations for
many Na concentrations.

\acknowledgments We thank J.~Bobroff, F.~Bert and P.~Mendels for performing
the $\mu$SR measurements on the paramagnetic phases and for helpful
discussions, as well as G.Kotliar, I.~Mazin, F.~Rullier-Albenque and
D.~Singh for their stimulating interest. We acknowledge financial support
from the ANR (NT05-4-41913), INTAS (04-83-3891) and RFBR (06-02-17197).

\end{document}